\documentclass[final,5p,times,twocolumn]{elsarticle}

\usepackage{color}
\usepackage[usenames,dvipsnames,svgnames,table]{xcolor}
\usepackage{hyperref}
\hypersetup{
     colorlinks   = true,
     citecolor    = gray
}

\biboptions{compress}

\journal{New Biotechnology: $\mathrm{November}$ $\mathrm{2009}$, Published: $\mathrm{July}$ $\mathrm{2010}$,  DOI: $\mathrm{\href{http://dx.doi.org/10.1016/j.nbt.2010.02.016}{10.1016/j.nbt.2010.02.016}}$}

\begin{document}

\begin{frontmatter}



\title{MicroRNAs -- targeting and target prediction}


\author[addr1]{Takay Saito}
\author[addr1,addr2]{P\aa{}l S\ae{}trom \corref{cor1}}
\ead{pal.satrom@ntnu.no}
\cortext[cor1]{Corresponding author}

\address[addr1]{Department of Cancer Research and Molecular Medicine, Norwegian University of Science and Technology, NO-7489 Trondheim, Norway}
\address[addr2]{Department of Computer and Information Science, Norwegian University of Science and Technology, NO-7491 Trondheim, Norway}

\begin{abstract}
MicroRNAs (miRNAs) are a class of small noncoding RNAs that can regulate many genes by base pairing to sites in mRNAs. The functionality of miRNAs overlaps that of short interfering RNAs (siRNAs), and many features of miRNA targeting have been revealed experimentally by studying miRNA-mimicking siRNAs. This review outlines the features associated with animal miRNA targeting and describes currently available prediction tools.
\end{abstract}




\end{frontmatter}


\section{Introduction}
\label{S:1}

MicroRNAs (miRNAs) were identified as a large sub-class of ncRNAs in 2001. Since then, an increasing number of studies have firmly established miRNAs’ importance in gene regulation in general and animal development and disease in particular \cite{Ambros2004, Soifer2007, Croce2009, Stefani2008, Bartel2004}. miRNAs regulate protein-coding genes post-transcriptionally by guiding a protein complex known as the RNA-induced silencing complex (RISC) to messenger RNAs (mRNAs) with partial complementarity to the miRNA \cite{Carthew2009}. Through mechanisms not completely understood, RISC then inhibits protein translation and causes mRNA degradation \cite{Bartel2009, Filipowicz2008}. Current estimates indicate that miRNAs regulate at least 60\% of the human protein-coding genes through this post-transcriptional gene silencing (PTGS) \cite{Friedman2009}.

Incorporated into RISC, miRNAs are functionally equivalent to short interfering RNAs (siRNAs) \cite{Hamilton1999, Zamore2000}. The main difference between these RNAs is that miRNAs are processed from imperfect hairpin structures, whereas siRNAs are processed from long double-stranded RNAs \cite{Kim2009, Winter2009}. Moreover, animal miRNAs typically target imperfect sites, whereas siRNAs target sites with near-perfect complementarity. SiRNAs do target imperfect sites as well, however, and this miRNA-like targeting is the major source of siRNA off-target effects \cite{Birmingham2006, Jackson2006, Jackson2003}.

The list of known miRNAs is large and increasing. Currently, the official miRNA database miRBase lists 721 human miRNAs (\url{http://www.mirbase.org}; Release 14) \cite{Griffiths-Jones2008}, but estimates indicate that the human genome contains more than 1000 miRNAs. As only a few regulatory targets are known, predicting and validating miRNA targets is one of the major hurdles in understanding miRNA biology. Here, we review the features important for miRNA targeting and the bioinformatics tools available for predicting miRNA targets.

\section{miRNA target features}

Identifying miRNA targets in animals has been very challenging. This is mainly because the limited complementarity between miRNAs and their targets, which might lead to the finding of hundreds of potential miRNA targets per miRNA. Therefore, many studies have been conducted, both experimentally and computationally, to reveal more efficient approaches for miRNA target recognition. We have divided the miRNA target features reported in these studies into six categories, miRNA:mRNA pairing, site location, conservation, site accessibility, multiple sites and expression profile.

\begin{figure}[h]
\centering\includegraphics[width=8.5cm]{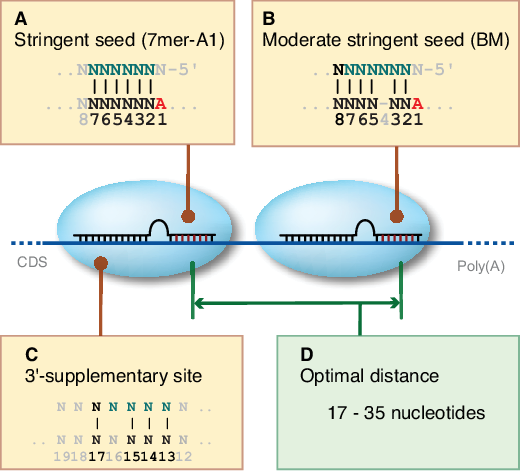}
\caption{Types of miRNA target sites and multiple sites. \textbf{(a)} Stringent-seed site. 7mer-A1 is shown as an example. Vertical lines indicate Watson-Crick paring. \textbf{(b)} Moderate-stringent-seed site. BM is shown as an example. \textbf{(c)} 3$'$-supplementary site. More than 3-4 nucleotide paring required. \textbf{(d)} Optimal distance of two miRNA target sites.}
\label{fig:fig1}
\end{figure}

\subsection{miRNA:mRNA pairing: `Seed site' is the most important feature for target recognition}

miRNA targets commonly have at least one region that has Watson-Crick pairing to the 5$'$ part of miRNA. This 5$'$ part, located at positions 2-7 from the 5$'$ end of miRNA, is known as the `seed', as RISC uses these positions as a nucleation signal for recognizing target mRNAs \cite{Stark2003, Rajewsky2004, Lewis2003}. The corresponding sites in mRNA are referred to as `seed sites'. A stringent-seed site has perfect Watson-Crick pairing and can be divided into four `seed' types -- 8mer, 7mer-m8, 7mer-A1 and 6mer -- depending on the combination of the nucleotide of position 1 and pairing at position 8 (Fig. \ref{fig:fig1}a). 8mer has both an adenine at position 1 of the target site and base pairing at position 8. 7mer-A1 has an adenine at position 1, while 7mer-m8 has base pairing at position 8. 6mer has neither an adenine at position 1 nor base pairing at position 8 \cite{Grimson2007}. An adenine on the target site corresponding to position 1 of miRNA is known to increase efficiency of target recognition \cite{Lewis2005}.

In addition to this stringent-seed matching, moderate-stringent-seed matching is also functional because RISC can tolerate small mismatches or G:U wobble pairing within the seed region (Fig. \ref{fig:fig1}b). This moderate-stringent-seed matching has five \lq{}seed\rq{} types: GUM, GUT, BM, BT and LP, defined regarding to the mismatch type. GUM has one G:U wobble and the uracil on the seed site of miRNA, whereas GUT has the uracil on the target site of mRNA. BM has one bulge and the mismatch is on the seed site, whereas BT has the mismatch on the target site. LP has only one loop \cite{Gaidatzis2007}. Furthermore, RISC can recognize offset sites that are located at positions 3-10. Offset sites can be either stringent or moderate-stringent-seed matching \cite{Hammell2008}.

Watson-Crick pairing in the 3$'$ part of miRNA is known to enhance the site recognition efficacy in miRNA targets that have seed pairing \cite{Grimson2007}. The preferable nucleotide number of matches in the 3$'$ part differs between the site that has stringent-seed pairing and the one that has moderate-stringent-seed pairing. Stringent-seeds require 3-4 matches in the positions 13-16, whereas moderate-stringent-seeds require 4-5 matches in the positions 13-19. Sites with this additional 3$'$ pairing are called 3$'$-supplementary (Fig. \ref{fig:fig1}c) and 3$'$-compensatory sites \cite{Bartel2009}.

It is difficult to measure the efficacy level of each seed type precisely, but several microarray and conservation enrichment studies have revealed hierarchies of relative efficacies. These hierarchies can be described as `Stringent seed' $>$ `Stringent seed in offset' $>$ `Moderate-stringent seed' $>$ `Moderate-stringent in offset'; 8mer $>$ 7mer-m8 $>$ 7mer-A1 $>$ 6mer in the stringent-seed types; and Bulge $>$ G:U wobble $>$ Loop in the moderate-stringent-seed types \cite{Bartel2009, Hammell2008}. Moreover, multiple sites are more efficient than single sites \cite{Saetrom2007}.

The advantages and disadvantages of using different set of seed types are that considering only stringent-seed types increases specificity but might miss many potential targets, whereas considering both stringent and moderate-stringent-seed types increase sensitivity but might also increase the number of false positives.

\subsection{Site location: most target sites reside within 3$'$ untranslated region (UTR) of target genes}
\label{S:2}

Several studies have reported that most target sites can be found in the 3$'$ UTR segment of the target genes, even though miRNA-loaded RISC in theory can bind any segment of mRNA. Target genes tend to have longer 3$'$ UTR, whereas ubiquitously expressed genes, such as house-keeping genes, have shorter 3$'$ UTR -- potentially to avoid being regulated by miRNAs \cite{Stark2005}. Target sites are not evenly distributed within 3$'$ UTR, but are located near both ends when the length of 3$'$ UTR is $\geq$2000 nucleotides. For shorter 3$'$ UTRs, sites tend to be near the stop codon \cite{Gaidatzis2007}. Sites are not located too close to the stop codon, however, but 15-20 nucleotides away from the stop codon \cite{Grimson2007}. In addition, some genes have alternative splicing in their 3$'$ UTR segments, especially genes with long 3$'$ UTRs. These genes might therefore have different potential target sites for alternatively spliced 3$'$ UTRs \cite{Majoros2007}. Finally, alternative polyadenylation sites can shorten 3$'$ UTRs and affect miRNA regulation \cite{Sandberg2008}.

Although functional miRNA sites are preferentially located in 3$'$ UTR, seed sites in the coding sequence (CDS) and 5$'$ UTR regions can also give downregulation \cite{Lytle2007, Kloosterman2004}. Why does RISC then appear to prefer the 3$'$ UTR? The most probable explanation is that RISC competes with other protein complexes, such as ribosomes in CDS and translation initiation complexes in 5$'$ UTR; see discussion in the following section `Multiple sites: cooperativity enhances site efficacy'. The 3$'$ UTR might simply be more accessible for long-term binding than the two other mRNA regions \cite{Bartel2004}.

Despite this general trend for 3$'$ UTR targeting, there are some notable exceptions. One recent study reported that many miRNAs preferentially target 5$'$ UTR sites with high complementarity to the miRNAs' 3$'$ end in a species-specific manner \cite{Lee2009}. The targets also showed signs of weaker interactions between the miRNA seed sequence and the 3$'$ UTR. The authors proposed that these sites represented a new miRNA target class called `miBridge', in which one miRNA simultaneously interacts with a seed pairing site in 3$'$ UTR and a 3$'$ pairing site in 5$'$ UTR. The molecular mechanisms behind and the biological extent of these miBridge targets are still unknown, however.

Most miRNA target prediction studies only focus on the 3$'$ UTR, which results in that all the available data are biased toward 3$'$ UTR. Moreover, few studies consider alternative splicing or polyadenylation because of shortcomings in current annotations. As transcript usage often depends on cellular context -- for example, whether the cell is proliferating or terminally differentiated -- future tools for miRNA target analyses should probably use available information about cellular state to increase prediction performance.

\subsection{Conservation: miRNAs and their targets are conserved among related species}

miRNA families are comprised of miRNAs that have the same seed site, and are well conserved among related species. In addition, miRNA families have targets that are conserved among related species \cite{Friedman2009}. There are also species-specific miRNAs and targets, and one study showed that about 30\% of the experimentally validated target genes might not be well conserved \cite{Sethupathy2006}.

siRNA off-target effects occur no matter whether the site is conserved or not \cite{Burchard2009}, therefore searching for all potential target sequences without considering evolutionally conservational might increase siRNA off-target detection efficacy.

Applying a filter that requires predicted target sites to be conserved can decrease the false-positive rate, but such a filter is effective only for conserved miRNAs. It is important to identify targets both with and without conservation -- especially when species-specific miRNAs or siRNA off-targets is of interest.

\subsection{Site accessibility: mRNA secondary structure affects site accessibility}

The mRNA secondary structure is very important for miRNA targeting. An effective miRNA:mRNA interaction needs an open structure on the target site to begin the hybridization reaction. After binding, RISC can disrupt the secondary structure on the site to elongate hybridization \cite{Long2007, Kertesz2007}. Minimum free energy is usually used to estimate the secondary structure and RNA hybridization, but the amount of A:Us surrounding the site can also be used to estimate the site accessibility. Effective target sites often have A:U rich context in approximately 30 nucleotides upstream and downstream from the seed matching region of the target site \cite{Grimson2007}.

Calculating the minimum free energy of accessibility and hybridization with the mRNA secondary structure requires analyzing different mRNA folding patterns. This requires enormous amounts of computing power, as finding the most stable RNA structure is a computational problem that scales with the cube of the length of the RNA sequence \cite{Eddy2004}. Hence, finding hybridization sites in long 3$'$ UTRs tends to be time consuming. Moreover, the current thermodynamic models used in RNA secondary structure prediction algorithms are only 90-95\% accurate, which results in that the algorithms tend to have only 50-70\% of the base pairs correct \cite{Eddy2004}. Thus, despite being theoretically sound, calculating site accessibility has limited practical value when predicting miRNA target sites; heuristics that are easy to compute, such as local A:U context, have similar performance.

\subsection{Multiple sites: cooperativity enhances site efficacy}

Strong miRNA targets tend to have multiple target sites instead of one single site \cite{John2004}. Considering the number of putative miRNA sites per mRNA can therefore significantly enhance target prediction.

Although the general effect of multiple sites appears to be additive, miRNA targeting can also be synergistic. Our previous study showed that two target sites within optimal distance enhance target site efficacy. The preferable optimal length is between 17 and 35 nucleotides, but the length between 14 and 46 nucleotides also enhances the efficacy (Fig. \ref{fig:fig1}d). This co-operability is functional between the same miRNAs as well as two different miRNAs \cite{Saetrom2007}. Multiple sites involving more than two sites can also contribute to the enhancement \cite{Doench2004}.

The exact mechanism underlying the synergism remains unknown. As translational suppression is a relatively slow process compared with RISC’s catalytic cleavage \cite{Haley2004}, however, multiple RISC complexes bound at closely spaced target sites might cooperatively stabilize each other at the sites or possibly accelerate the regulatory process. This could explain why miRNAs prefer targets in 3$'$ UTRs, as ribosomes would displace RISC from sites in CDS before RISC could effect translational suppression. Indeed, a cluster of rare codons that stall the ribosome can, when placed in front of a nonfunctional miRNA site in CDS, change the site to a functional site \cite{Gu2009}. Moreover, the genes that currently have verified miRNA target sites in CDS tend to have either one very strong target site \cite{Duursma2008, Elcheva2009} or multiple, closely spaced sites \cite{Tay2008, Forman2008}.

\subsection{Expression profile: miRNA:mRNA pairs are negatively correlated in expression profiles}

One miRNA can potentially regulate many genes; therefore, expression profiles of mRNAs might vary substantially depending on the miRNA expression levels. Many miRNAs are also expressed differently in different tissues. Consequently, if negatively correlated expression levels of a miRNA:mRNA pair are detected across different tissue profiles, the mRNA of the pair is probably targeted by the miRNA \cite{Rajewsky2006, Lim2005}. Filtering putative targets based on expression profile correlations is an effective approach to reduce the false-positive rate. Although the majority of miRNA targets appear to be regulated both at the mRNA and protein level, some targets only show an effect at the protein level, however \cite{Selbach2008, Baek2008}. Researchers should therefore be aware that such filtering will exclude potential targets.

\section{Target prediction tools}
\label{S:3}

Many target prediction tools have been developed (Table \ref{table:table1}), but the types of methods applied, the miRNA and mRNA sequences used and the output prediction data and performance evaluation vary widely between tools. Direct comparison of prediction performance among tools is not straight forward because the set of predicted target genes from different tools do not overlap well. What is clear, however, is that conventional tools with simple stringent-seed search are prone to high false-positive rates. Therefore, most tools are designed to reduce the false-positive rate and maximize the accuracy at the same time. We have compared the currently available tools based on the target features the tools use in their predictions (Table \ref{table:table1}), and the tools’ availability (Table \ref{table:table2}). Availability is especially important for researchers that are using their own miRNA or mRNA annotations, or are working in a nonstandard species. In these cases, only tools that can be downloaded or allows the user to input own miRNAs and mRNAs can be used.

\begin{table*}[!ht]
\caption{List of miRNA target prediction tools}
\begin{tabular*}{\textwidth}{@{\extracolsep{\fill} } l l l l l l l l}
\hline
\textbf{Tool} & \textbf{Pair$^\mathrm{a}$} & \textbf{Site$^\mathrm{b}$} & \textbf{Consv$^\mathrm{c}$} & \textbf{Access$^\mathrm{d}$} & \textbf{Multi$^\mathrm{e}$} & \textbf{Expr$^\mathrm{f}$} & \textbf{Refs} \\
\hline
\textbf{TargetScan} & $\circ$ & $\bullet$ & $\bullet$ & $\circ$ & $\circ$ &  & \cite{Friedman2009, Grimson2007, Lewis2005} \\
\textbf{PicTar}  & $\bullet$ &  & $\circ$ & $\bullet$ & $\bullet$ &  & \cite{Chen2006, Grun2005, Krek2005, Lall2006} \\
\textbf{miRanda}  & $\bullet$ &  & $\circ$ & $\bullet$ & $\circ$ &  & \cite{John2004, Betel2008} \\
\textbf{MicroCosm Targets} & $\bullet$ &  & $\circ$ & $\bullet$ & $\circ$ &  & \cite{Griffiths-Jones2008, Enright2003, Griffiths-Jones2006} \\
\textbf{RNAhybrid} & $\bullet$ &  &  & $\bullet$ &  &  & \cite{Kruger2006, Rehmsmeier2004} \\
\textbf{PITA} & $\bullet$ &  & $\bullet$ & $\bullet$ & $\circ$ &  & \cite{Kertesz2007} \\
\textbf{STarMir} & $\bullet$ &  &  & $\bullet$ &  &  & \cite{Long2007} \\
\textbf{Rajewsky \& Socci} & $\bullet$ &  &  & $\bullet$ &  &  & \cite{Rajewsky2004} \\
\textbf{Robins} & $\bullet$ &  &  & $\bullet$ & $\circ$ &  & \cite{Robins2005} \\
\textbf{mirWIP} & $\bullet$ &  & $\circ$ & $\bullet$ & $\circ$ & $\bullet$ & \cite{Hammell2008} \\
\textbf{MicroInspector} & $\bullet$ &  &  & $\bullet$ &  &  & \cite{Rusinov2005} \\
\textbf{MicroTar} & $\bullet$ &  &  & $\bullet$ &  &  & \cite{Thadani2006} \\
\textbf{MirTarget2} & $\circ$ & $\bullet$ & $\bullet$ & $\bullet$ &  &  & \cite{Wang2008} \\
\textbf{miTarget} & $\bullet$ &  &  & $\bullet$ &  &  & \cite{Kim2006} \\
\textbf{TargetMiner} & $\bullet$ &  & $\circ$ & $\bullet$ &  & $\bullet$ & \cite{Bandyopadhyay2009} \\
\textbf{EIMMo} & $\bullet$ &  & $\circ$ &  & $\circ$ &  & \cite{Gaidatzis2007} \\
\textbf{NbmiRTar} & $\bullet$ &  & $\circ$ & $\bullet$ &  &  & \cite{Yousef2007} \\
\textbf{TargetBoost} & $\bullet$ &  &  &  &  &  & \cite{Saetrom2005} \\
\textbf{RNA22} & $\bullet$ &  & $\circ$ & $\bullet$ & $\bullet$ &  & \cite{Miranda2006} \\
\textbf{TargetRank} & $\circ$ &  & $\bullet$ & $\circ$ &  &  & \cite{Nielsen2007} \\
\textbf{EMBL} & $\bullet$ &  & $\circ$ & $\bullet$ & $\circ$ &  & \cite{Stark2003}\cite{Stark2005}\cite{Brennecke2005} \\
\textbf{MovingTarget} & $\bullet$ &  & $\circ$ & $\bullet$ & $\circ$ &  & \cite{Burgler2005} \\
\textbf{DIANA-microT} & $\bullet$ &  & $\circ$ & $\bullet$ &  &  & \cite{Kiriakidou2004} \\
\textbf{HOCTAR} & $\bullet$ &  & $\circ$ & $\bullet$ &  & $\bullet$ & \cite{Gennarino2009} \\
\textbf{Stanhope} &  &  &  &  &  & $\bullet$ & \cite{Stanhope2009} \\
\textbf{GenMiR++} & $\circ$ &  & $\circ$ &  &  & $\bullet$ & \cite{Huang2007} \\
\textbf{HuMiTar} & $\bullet$ &  &  &  &  &  & \cite{Ruan2008} \\
\textbf{MirTif} & $\bullet$ &  &  &  &  &  & \cite{Yang2008} \\
\textbf{Yan et al.} & $\bullet$ &  & $\circ$ & $\bullet$ &  &  & \cite{Yan2007} \\
\textbf{Xie et al.} & $\circ$ &   & $\circ$ &  &  &  & \cite{Xie2005} \\
\hline
\end{tabular*}

\begin{flushleft}
$^\mathrm{a}$miRNA:mRNA pairing. $\bullet$: stringent seeds, $\circ$: moderately stringent seeds, Blank: seed sites not considered. \\
$^\mathrm{b}$Site location. $\bullet$: target positions considered, Blank: target positions not considered. \\
$^\mathrm{c}$Conservation. $\bullet$: with/without conservation filter, $\circ$: with conservation filter, Blank: conservation not considered. \\
$^\mathrm{d}$Site accessibility. $\bullet$: site accessibility with minimum free energy considered, $\circ$: A:U rich flanking considered, Blank: site accessibility not considered. \\
$^\mathrm{e}$Multiple sites. $\bullet$: multiple sites considered, $\circ$: the number of putative sites consided, Blank: multiple co-operability not considered. \\
$^\mathrm{f}$Expression profile. $\bullet$: expression profiles used, Blank: expression profiles not used.
\end{flushleft}
\label{table:table1}
\end{table*}

\begin{table*}[!ht]
\caption{Resource availability for miRNA target prediction tools}
\begin{tabular*}{\textwidth}{@{\extracolsep{\fill} } p{0.12\textwidth} p{0.07\textwidth} p{0.05\textwidth}  p{0.05\textwidth}  p{0.05\textwidth}  p{0.03\textwidth}  p{0.63\textwidth}  }
\hline
\textbf{Tool} & \textbf{Predicted species$^\mathrm{a}$} & \multicolumn{3}{c}{\textbf{Web access}}  &  \textbf{SW$^\mathrm{d}$} & \textbf{URL} \\
\cline{3-5}
 &  & \textbf{Online tool} & \textbf{Own miRNA$^\mathrm{b}$} & \textbf{Own mRNA$^\mathrm{c}$} &  & \\
\hline
\textbf{TargetScan} & 23 vertebrates, f, w & Yes & Yes & Yes & Yes & \href{http://www.targetscan.org}{http://www.targetscan.org}  \\
\textbf{PicTar}  & v, m, f, w & Yes & No & No & No & \href{http://pictar.mdc-berlin.de}{http://pictar.mdc-berlin.de}  \\
\textbf{miRanda}  & h, m, r & Yes & No & No & Yes & \href{http://www.microrna.org}{http://www.microrna.org} \\
\textbf{MicroCosm Targets}  & 44 species & Yes & No & No & No & \href{http://www.ebi.ac.uk/enright-srv/microcosm/htdocs/targets/v5}{http://www.ebi.ac.uk/enright-srv/microcosm/htdocs/targets/v5} \\
\textbf{RNAhybrid} &  & No & No & No & Yes & \href{http://bibiserv.techfak.uni-bielefeld.de/rnahybrid}{http://bibiserv.techfak.uni-bielefeld.de/rnahybrid} \\
\textbf{PITA} & h, m, f, w & Yes & Yes & Yes & Yes & \href{http://genie.weizmann.ac.il/pubs/mir07}{http://genie.weizmann.ac.il/pubs/mir07} \\
\textbf{STarMir} &  & Yes & Yes & Yes & No & \href{http://sfold.wadsworth.org/starmir.pl}{http://sfold.wadsworth.org/starmir.pl} \\
\textbf{Rajewsky \& Socci} & f & No & No & No & No & \\
\textbf{Robins} & f, w & No & No & No & No & \\
\textbf{mirWIP} & w & Yes & No & No & Yes & \href{http://146.189.76.171/query.php}{http://146.189.76.171/query.php} \\
\textbf{MicroInspector} &  & Yes & Yes & Yes & No & \href{http://mirna.imbb.forth.gr/microinspector}{http://mirna.imbb.forth.gr/microinspector} \\
\textbf{MicroTar} &  & No & No & No & Yes & \href{http://tiger.dbs.nus.edu.sg/microtar}{http://tiger.dbs.nus.edu.sg/microtar} \\
\textbf{MirTarget2} & h, m, r, d, c & Yes & No & No & No & \href{http://mirdb.org}{http://mirdb.org} \\
\textbf{miTarget} &  & Yes & Yes & Yes & No & \href{http://cbit.snu.ac.kr/$\sim$miTarget}{http://cbit.snu.ac.kr/$\sim$miTarget} \\
\textbf{TargetMiner} & h & Yes & No & No & No & \href{http://www.isical.ac.in/$\sim$bioinfo\_miu}{http://www.isical.ac.in/$\sim$bioinfo\_miu} \\
\textbf{EIMMo} & h, m, f, z & Yes & No & Yes & No & \href{http://www.mirz.unibas.ch/ElMMo2}{http://www.mirz.unibas.ch/ElMMo2} \\
\textbf{NBmiRTar} &  & Yes & Yes & Yes & No & \href{http://wotan.wistar.upenn.edu/NBmiRTar}{http://wotan.wistar.upenn.edu/NBmiRTar} \\
\textbf{TargetBoost} & w & Yes & Yes & No & No & \href{https://demo1.interagon.com/targetboost}{https://demo1.interagon.com/targetboost} \\
\textbf{RNA22} &  & Yes & Yes & Yes & No & \href{http://cbcsrv.watson.ibm.com/rna22.html}{http://cbcsrv.watson.ibm.com/rna22.html} \\
\textbf{TargetRank} & h, m & Yes & No & No & No & \href{http://hollywood.mit.edu/targetrank}{http://hollywood.mit.edu/targetrank} \\
\textbf{EMBL} & f & Yes & No & No & No & \href{http://www.russell.embl-heidelberg.de/miRNAs}{http://www.russell.embl-heidelberg.de/miRNAs} \\
\textbf{MovingTarget} & f & No & No & No & No & \\
\textbf{DIANA-microT} &  & Yes & Yes & Yes & No & \href{http://diana.pcbi.upenn.edu/cgi-bin/micro\_t.cgi}{http://diana.pcbi.upenn.edu/cgi-bin/micro\_t.cgi} \\
\textbf{HOCTAR} & h & Yes & No & No & No & \href{http://hoctar.tigem.it}{http://hoctar.tigem.it} \\
\textbf{Stanhope} & h & No & No & No & No & \\
\textbf{GenMiR++} & h & No & No & No & Yes & \href{http://www.psi.toronto.edu/genmir/}{http://www.psi.toronto.edu/genmir/} \\
\textbf{HuMiTar} & h & No & No & No & No & \\
\textbf{MirTif} &  & Yes & Yes & Yes & No & \href{http://bsal.ym.edu.tw/mirtif}{http://bsal.ym.edu.tw/mirtif} \\
\textbf{Yan et al.} & h & No & No & No & No & \\
\textbf{Xie et al.} & h, m, r, d & No & No & No & No & \\
\hline
\end{tabular*}

\begin{flushleft}

$^\mathrm{a}$Both species of pre-computed prediction and the species available on the web tool are listed. Letters indicate the species: fly (f), worm (w), human (h), mouse (m), rat (r), chicken (c), zebra fish (z), and dog (d). Cells are left empty when no information is available. \\
$^\mathrm{b}$Yes/No indicate whether own miRNA sequences can be used on the web interface or not. \\
$^\mathrm{c}$Yes/No indicate whether own mRNA sequences can be used on the web interface or not. \\
$^\mathrm{d}$SW: Software availability (executable or source code).

\end{flushleft}
\label{table:table2}
\end{table*}

Most tools rely on either one or a combination of seed matching, site accessibility and evolutionary conservation features, although some recently developed tools use expression profiles. No tools have successfully incorporated some of the important features, such as optimal distances of multiple miRNA sites or supplementary sites in CDS and 5$'$ UTR.

TargetScan \cite{Friedman2009, Grimson2007, Lewis2005}, PicTar \cite{Chen2006, Grun2005, Krek2005, Lall2006} , miRanda \cite{John2004, Betel2008}, RNAhybrid \cite{Kruger2006, Rehmsmeier2004} and PITA \cite{Kertesz2007} have been frequently used for performance comparisons or as preprocessors for other tools to obtain initial putative target sites. Of the five, TargetScan often shows the best performance in comparisons. TargetScan considers only stringent seeds, however, and therefore ignores many potential targets.

\section{Summary}
\label{S:4}

Finding true functional miRNA targets is still challenging even though many biological features of miRNA targeting have been revealed experimentally and computationally. Building a model with more features might achieve higher accuracy and enhance site recognition efficacy, but its implementation might also become more complex. None of the existing prediction tools has been able to incorporate all currently known features. We expect that a new approach that can combine the features from the six categories we have shown will significantly improve computational miRNA target prediction.

Another important problem that has hardly been addressed is predicting target interactions between different miRNAs. Different miRNAs can cooperatively regulate individual targets, but miRNA expression signatures differ between cell types and cellular conditions. Determining how varying miRNA expression affects target regulation in cancerous versus normal cells, for example, will therefore be a major problem in the coming years.




\bibliographystyle{model1-num-names}
\bibliography{mirna}







\end{document}